\documentclass[aps,prb,twocolumn,floatfix,showpacs,amsmath,amssymb]{revtex4-1}

\usepackage{graphicx}
\usepackage{dcolumn}
\usepackage{hyperref}

\begin{document}

\title{Extending spin ice concepts to another geometry: the artificial triangular spin ice}

\author{L.A.S. M\'{o}l}
\email{lucasmol@ufv.br}
\author{A.R. Pereira}
\email{apereira@ufv.br}
\author{W.A. Moura-Melo}
\email{winder@ufv.br} \affiliation{Departamento de F\'isica,
Universidade Federal de
Vi\c cosa, Vi\c cosa, 36570-000, Minas Gerais, Brazil \\
}

\begin{abstract}
In this work we propose and study a realization of an artificial spin ice-like system in a triangular
geometry, which unlike square and kagome artificial spin ice, is not based on any
real material. At each vertex of the lattice, the ``ice-like rule''
dictates that three spins must point inward while the other three must point outward. We have
studied the system's ground-state and the lowest energy excitations as well as the thermodynamic
properties of the system. Our results show that, despite fundamental differences in the vertices topologies
as compared to the artificial square spin ice, in the triangular array the lowest energy excitations
also behave as a kind of Nambu monopoles (two opposite monopoles connected by an energetic string).
Indeed, our results suggest that the monopoles charge intensity may have a universal value while the
string tension could be tuned by changing the system's geometry, probably allowing the design
of systems with different string tensions. Our Monte Carlo findings suggest a phase transition
in the Ising universality class where the mean
distance between monopoles and anti-monopoles increases considerably at the critical temperature.
The differences on the vertices topologies seem to facilitate the experimental achievement of the
system's ground-state, thereby allowing a more detailed experimental study of the system's properties.
\end{abstract}

\maketitle

\section{Introduction}

Geometrical frustration and fractionalization are key concepts in modern
condensed matter theory. While the former is related to the impossibility
of simultaneously minimize the energy for all constituents of a system due
to geometrical constraints, the latter is related to the appearance of
collective excitations that carry only a fraction of the elementary
constituents properties. In some systems these phenomena are closely related,
as is the case of spin ice materials (Dy$_2$Ti$_2$O$_7$ and Ho$_2$Ti$_2$O$_7$)
\cite{Castelnovo08,Fennell09,Morris09,Bramwell09,Kadowaki09,Jaubert09,Giblin11}.
There, due to a geometrical frustration, the energy is
minimized by the appearance of the two-in/two-out ice rule, where at each vertex
of a lattice two spins point inward and two outward. Violations of
this rule are viewed thus as a fraction of a dipole~\cite{Castelnovo08}, since these collective
excitations behave as magnetic monopoles, constituting the first example of
fractionalization in three-dimensional ($3d$) materials.

With the aid of modern experimental techniques,
mainly the capability to construct and manipulate nanostructured systems, artificial arrays with
properties very similar to the spin ice material were recently built~\cite{Wang06,Li10,Zabel09,Ladak10,Mengotti10}.
In particular, in an artificial frustrated two-dimensional ($2d$) square array that mimic the
spin ice behavior, the excitations above the ground-state are viewed as a kind of Nambu magnetic
monopoles~\cite{Nambu77,Mol09,Mol10,Morgan10,Moller09}, since the end points of the energetic
string behave like particles with magnetic charge, which leads to a Coulomb interaction
(we remark that the Coulomb is a particular case of the Yukawa potential present in the Nambu calculations).
Therefore, there is, of course, a huge interest in accessing the ground-state of the artificial
square spin ice (ASSI) to test theoretical predictions concerning the appearance and behavior
of monopoles excitations \cite{Moller06,Silva12,Budrikis10,Kapaklis11}. However demagnetization protocols~\cite{Ke08,Nisoli07,Nisoli10} used so far
were not able to drive the ASSI to its lowest energy state.
Morgan {\it et al}~\cite{Morgan10} successfully achieved what seems to be the
thermalization of the ASSI during fabrication, although, they could not obtain a single ground-state domain.

In this work we investigate if the same fractionalization phenomenon manifested in the ASSI
is also present in another artificial setting. We consider here a particular realization
of an artificial spin ice in a triangular geometry. Unlike the square and kagome spin ice, our proposal
 is not based in any real material
and were motivated by inquiring what changes arise when the geometry of an artificially frustrated
magnetic system is not the usual for typical spin ices. By doing this we have found that in the triangular
geometry the ground-state is much likely to be easily obtained experimentally, allowing the experimental
investigation of the existence and behavior of collective excitations above the ground-state. In addition, we have found
that the artificial triangular system also has collective excitations that can be described by a kind of
Nambu monopole. As expected for a system which can be described by the same kind of excitation present in
the ASSI, many similarities with ASSI were found. On the other hand, there are fundamental
differences between the ASSI and the artificial triangular spin ice (ATSI), specially in what concerns the vertices'
topologies. The existence of monopole-like excitations in the ATSI may also stimulate the searching for other
crystalline 3D materials where magnetic monopoles could appear.

\section{The Artificial Triangular Spin Ice (ATSI)}

\begin{figure}
\begin{center}
\begin{tabular}{cc}
\includegraphics[scale=0.18]{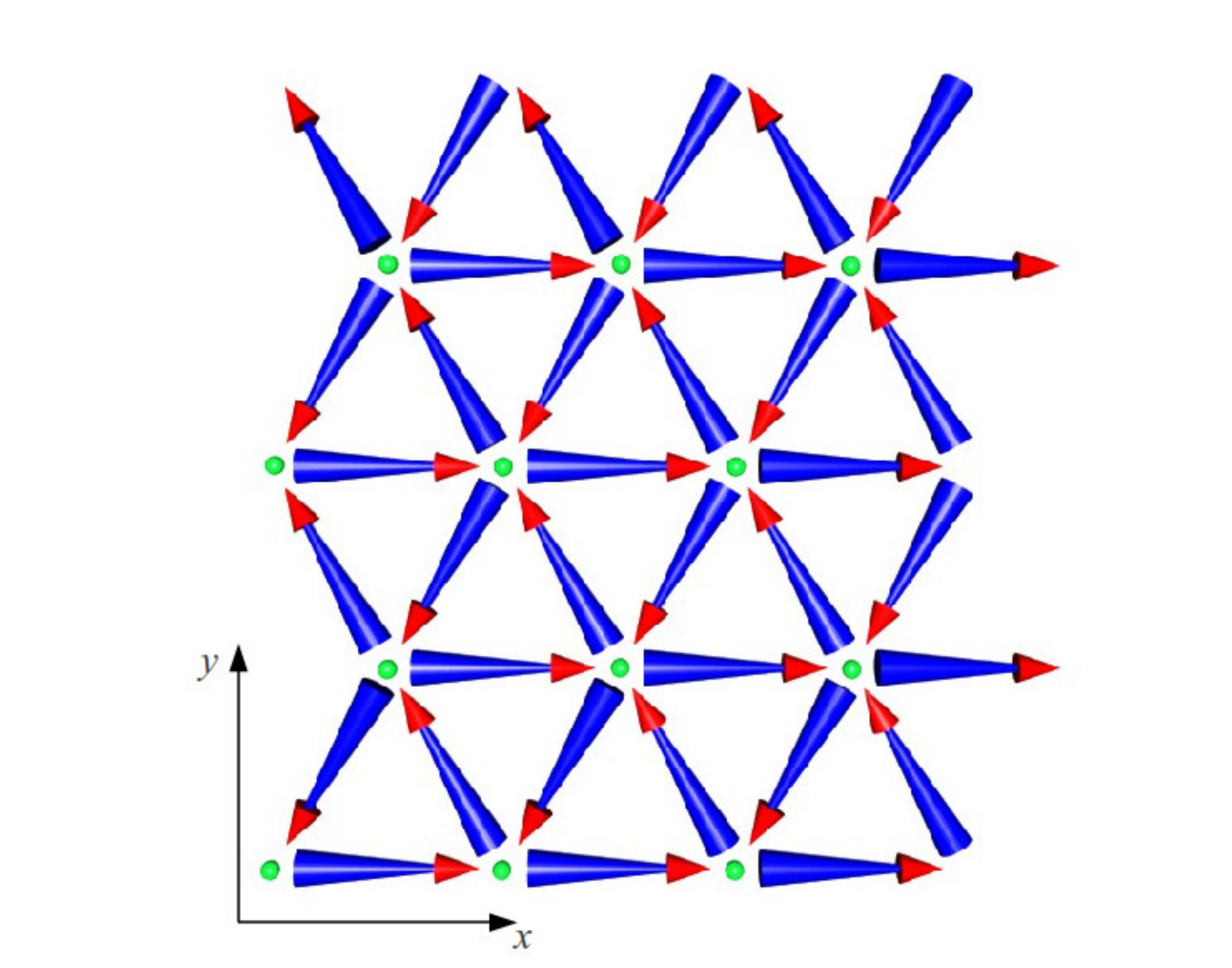} & \includegraphics[scale=0.17]{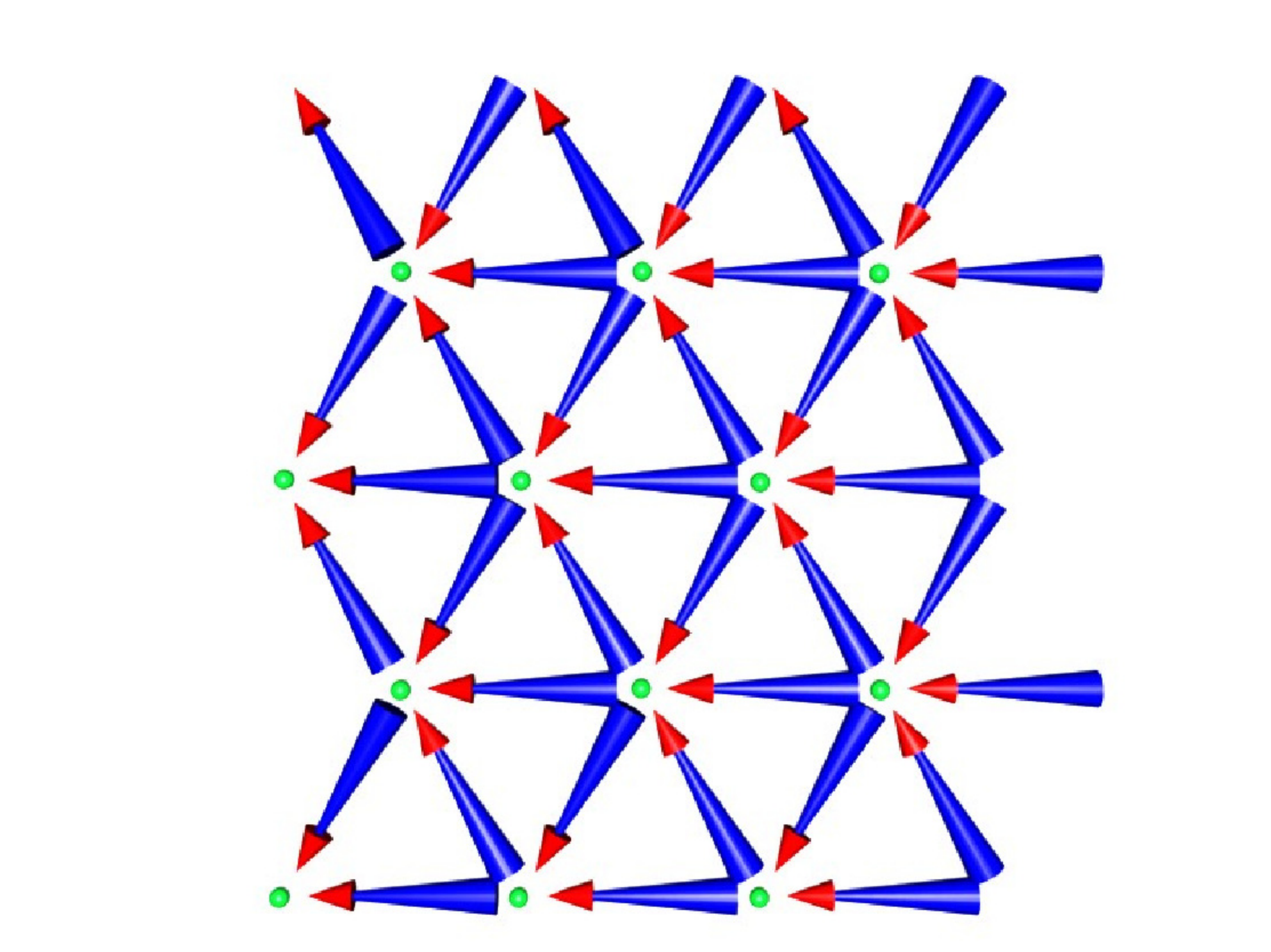}
\end{tabular}
\end{center}
\caption{ \label{array} (Color online) Left: A lattice with $L=3$ in its ground-state. The array proposed in
this work has six spin in each vertex (small dots). The ice rule is now three in/three out. Right: By applying
an strong magnetic field in the $-x$ direction the state shown here should be achieved and by reversing
the applied field the system should be driven to the ground-state shown on the left side of the figure.
}
\end{figure}

The study of spin ice-like systems is somehow restricted, until now, to two kinds of vertices:
vertices with three spins, as in the kagome and brickwork lattices, and vertices with four spins
as in the natural spin ice materials (Dy$_2$Ti$_2$O$_7$ and Ho$_2$Ti$_2$O$_7$) and in the
artificial square spin ice (ASSI). We may therefore ask what happens if we have a lattice
with six spins per vertex instead of three or four. This is the basic question that motivated this study.
To answer it we have considered an array of elongated magnetic nanoparticles having
a single domain pattern, as those used to build the ASSI (Permalloy nanoislands), placed in a geometry
such that the longest axis of each island points along the lines joining two vertices in a triangular lattice (see Fig.~\ref{array}).

We have modeled the proposed system by replacing the magnetic moment of each island by a point-like,
Ising-like dipole ($\vec{S}_i$) that is constrained to point along the line that joins two
vertices in the lattice, such that, the interactions are expressed as follows:
\begin{eqnarray}\label{HamiltonianSI}
H_{SI} &=& Da^{3} \sum_{i\neq j}\left[\frac{\vec{S}_{i}\cdot
\vec{S}_{j}}{r_{ij}^{3}} - \frac{3 (\vec{S}_{i}\cdot
\vec{r}_{ij})(\vec{S}_{j}\cdot \vec{r}_{ij})}{r_{ij}^{5}}\right],
\end{eqnarray}\\
where $D=\mu_{0}\mu^{2}/4\pi a^{3}$ is the dipolar interaction coupling constant, $a$ is the lattice spacing and
$\vec{r}_{ij}$ is a vector that connects islands $i$ and $j$.
In Fig.~\ref{array} we present a sketch of the system, where the vertices and the island's
magnetic moments are shown.

We start our analysis by considering a single vertex
and its six spins. It can be easily shown that it is energetically favorable
when the moments of a pair of islands are aligned, so that one is pointing
into the center of the vertex and the other is pointing
out, while it is energetically unfavorable when both moments are pointing inward
or both are pointing outward. Therefore the system is expected to be frustrated because for a particular
configuration obeying the ice-like rule (three-in/three-out),
from the $15$ possible pairs of islands in a vertex, only $9$ can minimize the energy.
For the sake of comparison we remember that in the square spin ice there are $6$ pairs at
each vertex and only $4$ can minimize the energy when the ice rule is considered.

\begin{figure}
\includegraphics[scale=0.22]{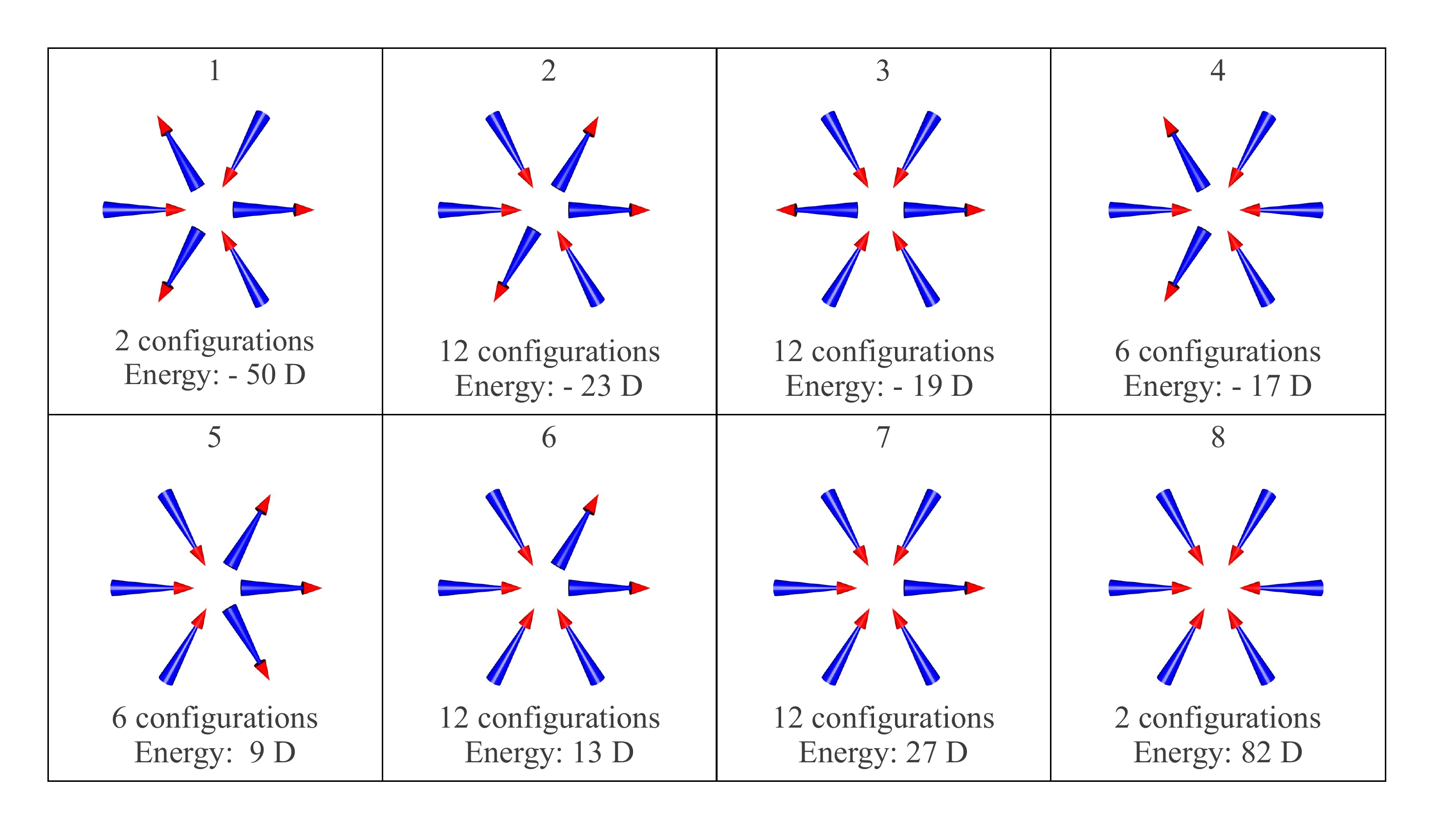}
\caption{ \label{Config}(Color online) The $64$ possible vertices, grouped by increasing energy (left to right)
in $8$ topologies. Vertices types 1, 2 and 5 satisfies the three-in/three-out ice rule, while vertices
types 3, 4 and 6 are single magnetic charges. Vertices type 7 and 8 are double and triple charged vertices
respectively.}
\end{figure}

Each vertex of the system has $64$ configurations distributed in $8$ different topologies
(let us recall that the square lattice has only $16$ configurations distributed
in $4$ topologies). In Fig.~\ref{Config}
we show these configurations and topologies, grouped by increasing energy.
Types $1,2$ and $5$ obey the three-in/three-out ``ice rule", and analogously to the
square lattice, they are energetically split. The remaining types of vertices shown
in Fig.~\ref{Config} possess monopole-like magnetic charges. These correspond
to the flip of one, two, or even three spins in a particular vertex.
For instance, types $3,4$
and $6$ are monopoles with single charges (with four-in/two-out or two-in/four-out configurations), while
types $7$ and $8$ are doubly (five-in/one-out or one-in/five-out) and triply
(six-in or six-out) charged monopoles respectively. There are, therefore, $3$
different classes of monopoles with single charge and only one class of doubly
and triply charged.

At this point, a fundamental difference between the ASSI and the ATSI appear:
while in the ASSI all vertices violating the ice-rule have more energy than
the vertices that satisfy it, for the ATSI, this is not true. In an energy
scale we have type 1 and type 2 vertices, which satisfies the ice-rule, followed by vertices types 3 and 4, which has four-in/two-out or four-out/two-in spins and of course does not satisfies the ice-rule. Indeed, vertices types 3 and 4 has less energy than type 5 vertices,
which has three spins in, three spins out. From this perspective
one should expect many differences in the collective behavior of the ATSI and ASSI,
for example, in the lowest energy excitations and in the thermodynamic properties;
but as will be shown soon they are much more similar than expected.

Going beyond the study of a single vertex we have checked by a simulated annealing process
(a Monte Carlo simulation where the system's temperature is slowly decreased~\cite{Mol09})
that the system's ground-state is indeed composed only by type 1 vertices as can be expected
by the vertices energies shown in Fig.~\ref{Config}. Details of the
Monte Carlo procedure, including lattice sizes, etc are given in section~\ref{termo}. A sketch of
the system in its ground-state is shown in Fig.~\ref{array}.

\section{Lowest energy excitations}

\begin{figure}
\begin{tabular}{cc}
\includegraphics[scale=0.15]{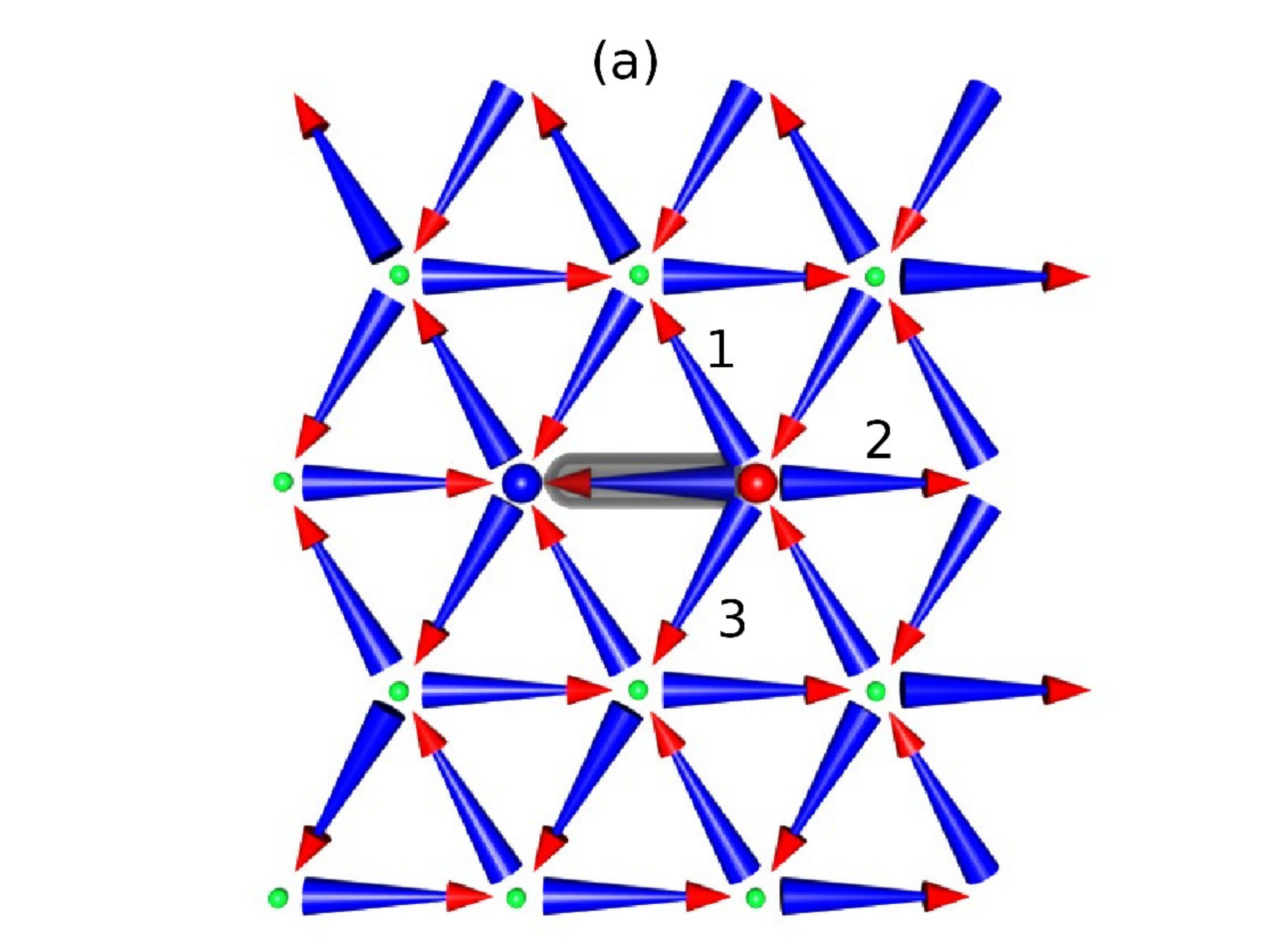}&\includegraphics[scale=0.15]{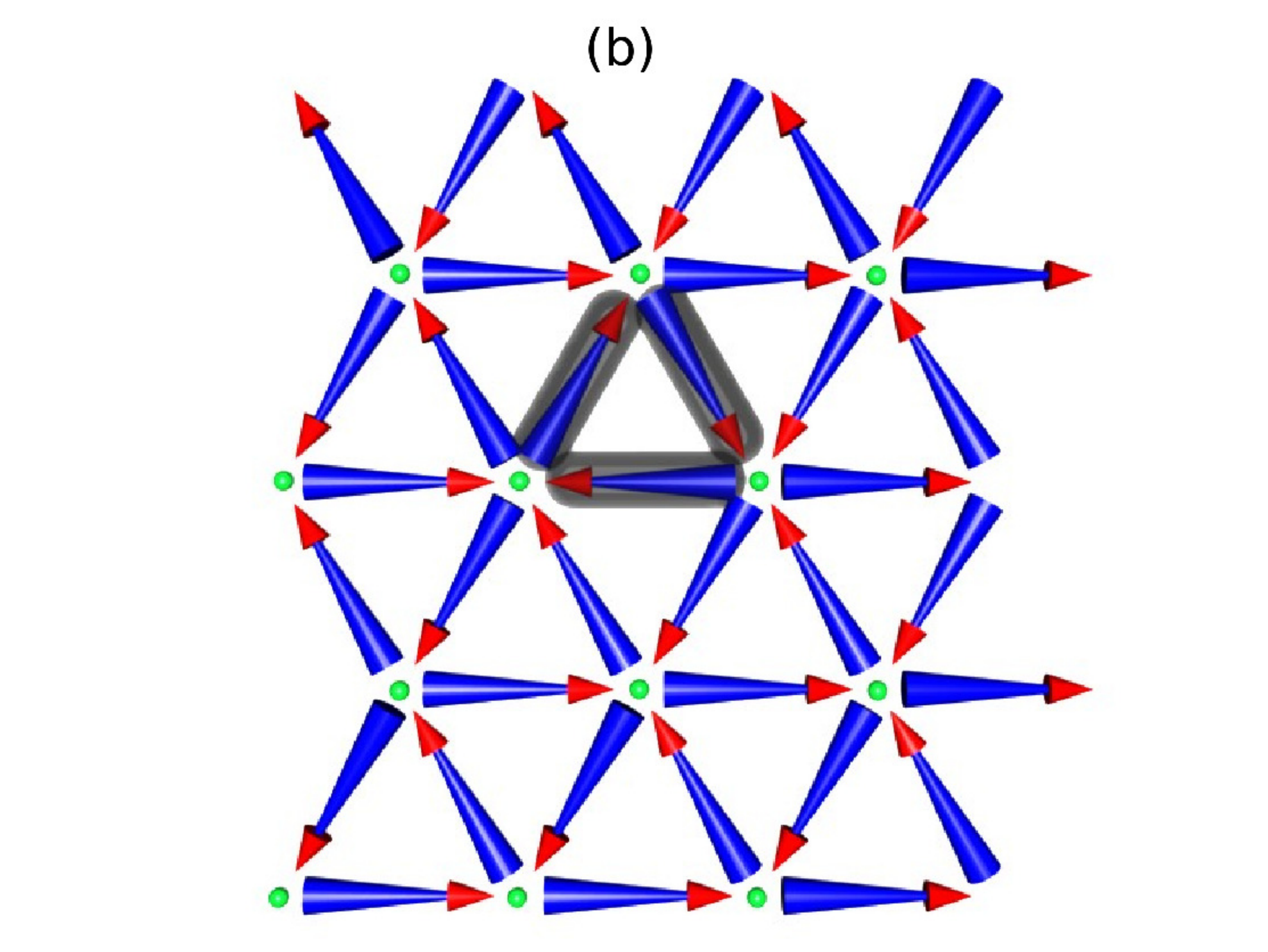}\\
\includegraphics[scale=0.15]{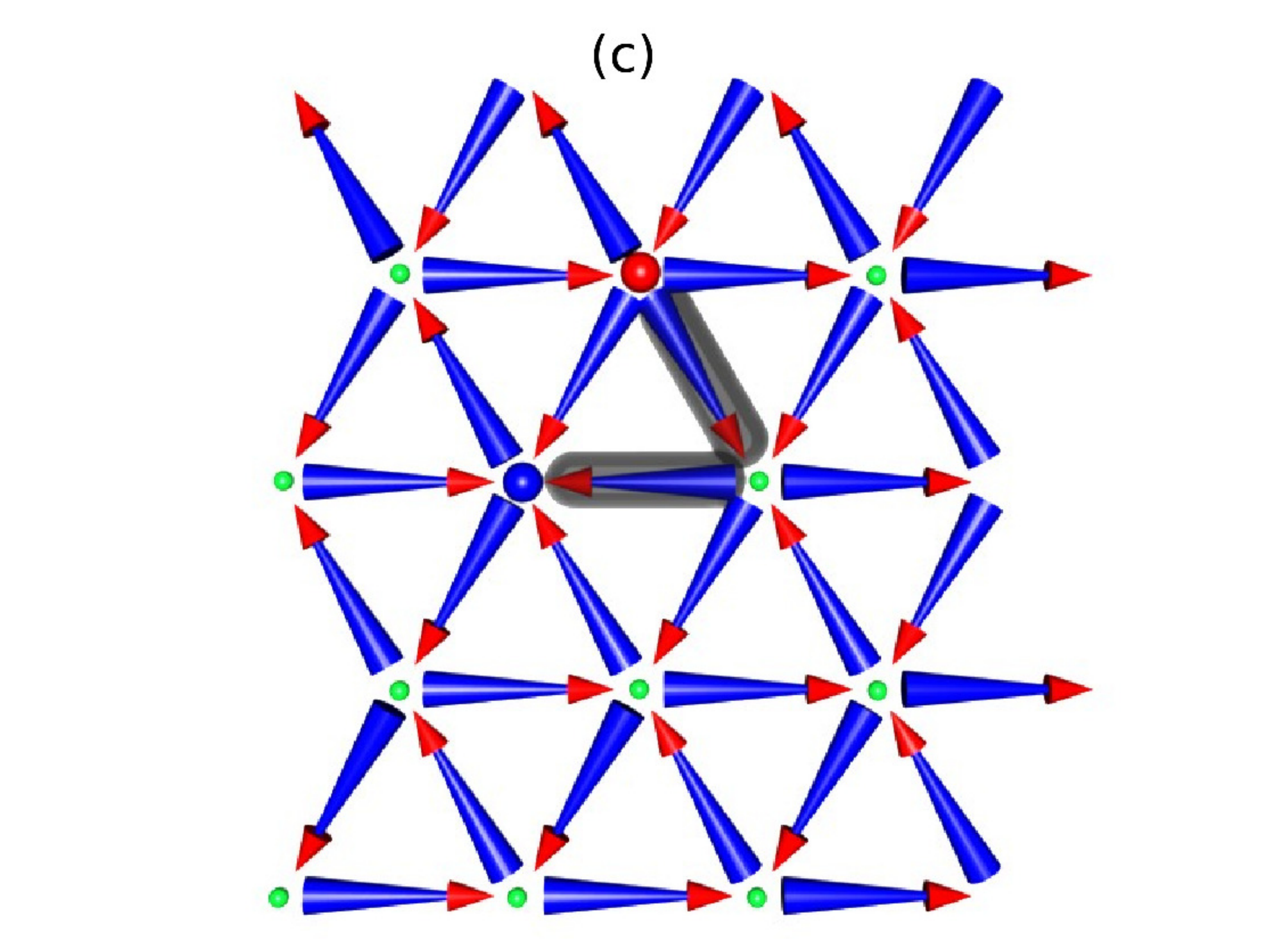}&\includegraphics[scale=0.15]{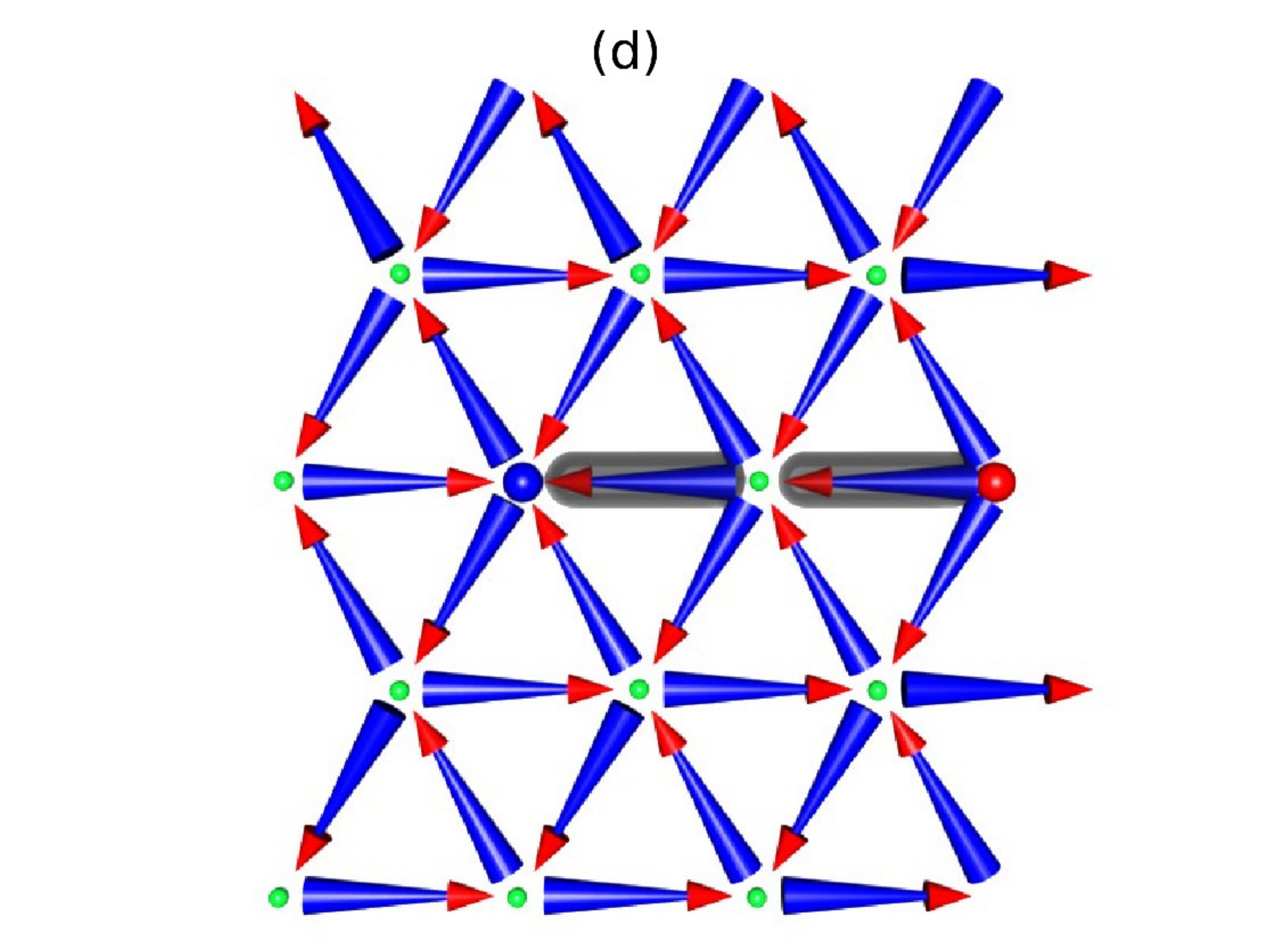}\\
\includegraphics[scale=0.15]{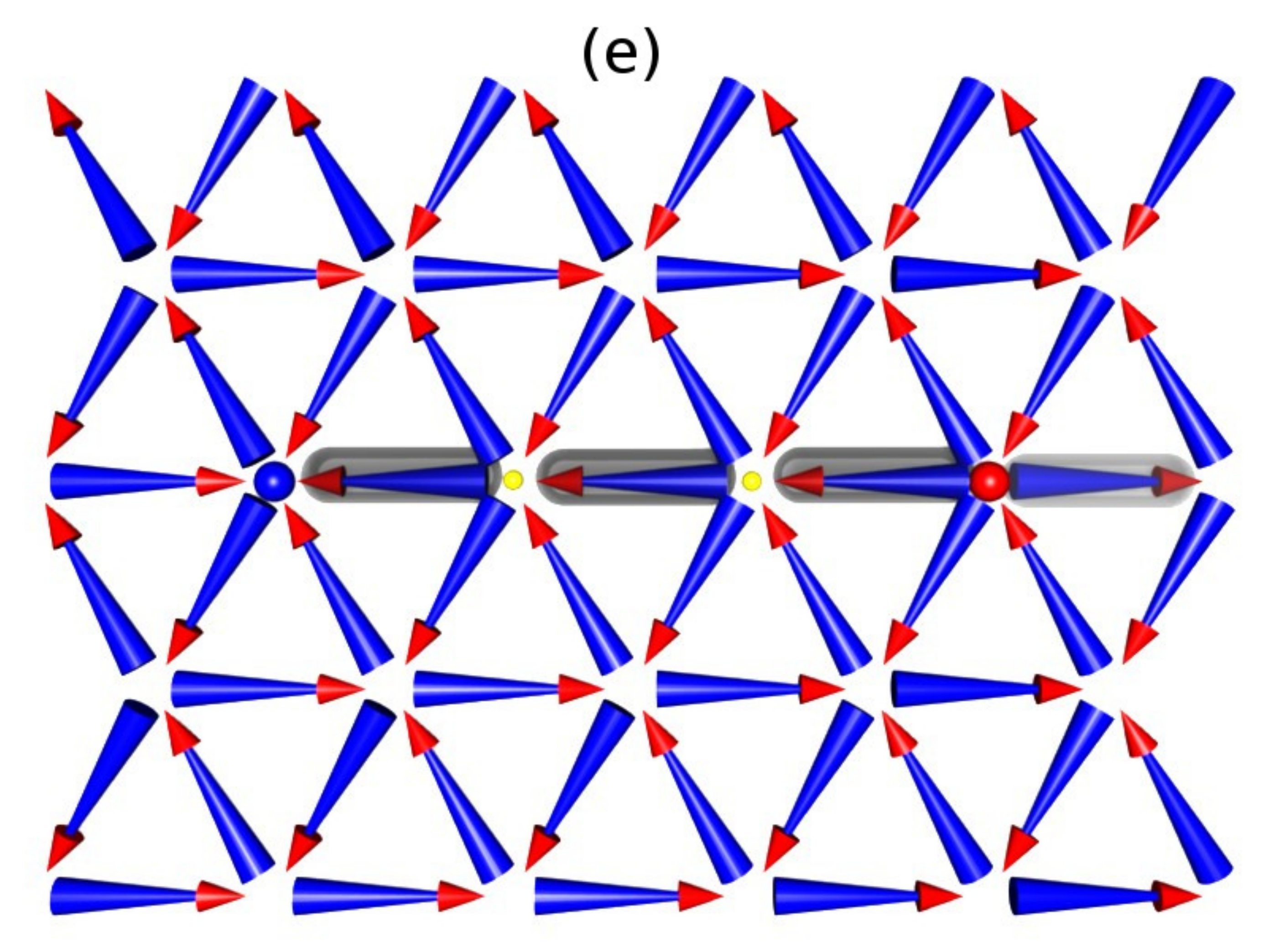}&\includegraphics[scale=0.15]{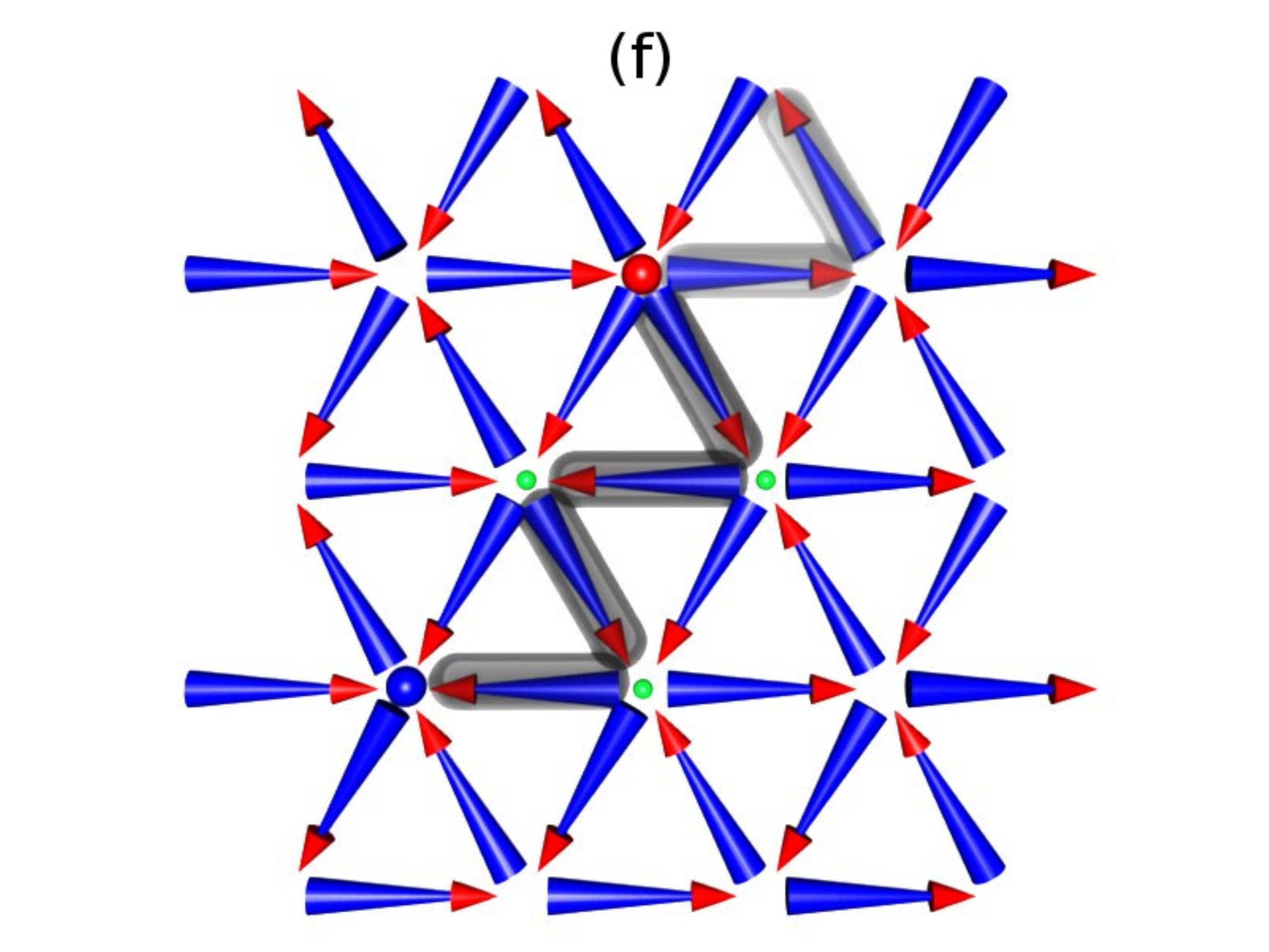}
\end{tabular}
\caption{ \label{low-energy} (Color online) (a) Flipping the spin marked in gray results in a monopole
anti-monopole pair (big dots). This is the lowest energy excitation of the system.
By flipping spins 1, 2 or 3 the configurations of (c) and (d) can be obtained.
(b) The second lowest energy state, which is obtained by flipping the marked spins.
(c) The third lowest energy state. By repeating this process a sawtooth like string can be formed.
(d) The fourth lowest energy state, obtained by flipping the marked spins.
(e) A linear string path, where spins marked in dark gray were flipped. By flipping the spin marked
in light gray the monopoles are separated by one more lattice spacing. Type 5 vertices that compose the string
are marked with small (yellow) dots.
(f) A sawtooth string, where spins marked in dark gray were flipped. By flipping the spins marked
in light gray the monopoles are separated by one more lattice spacing while the string length
increases by two lattice spacings. Type 2 vertices that compose the string
are marked with small (green) dots.}
\end{figure}

The lowest energy excitation that can be obtained above the ground-state is achieved by flipping one
spin, creating two type 4 vertices as shown in Fig~\ref{low-energy} (a). This process has an energy cost
of approximately $66D$. The second lowest energy state is obtained by flipping a plaquete
(Fig.~\ref{low-energy} b) and has an energy cost of approximately $77D$. In this case three type 2 vertices
are created.

Of particular interest is the case where after flipping one spin, which creates two type 4 vertices,
more spins are flipped (without further violation of the ice rule) in order to separate them.
As can be seen in Fig.~\ref{low-energy} (a), to move a excitation to the right side,
keeping the neutrality (three-in/three-out) in between, we have three options:
flip the spins 1, 2 or 3. By symmetry, its clear that flipping spins 1
or 3 has the same effect in the system and generates the configuration shown in Fig.~\ref{low-energy} (c),
where a type 2 vertex is created. On the other hand, if spin 2 in Fig.~\ref{low-energy} (a) is flipped,
the configuration shown in Fig.~\ref{low-energy} (d) is formed, which comprises the appearance of a type 5
vertex. As expected, the configuration in Fig.~\ref{low-energy} (c) has less energy than the configuration of
Fig.~\ref{low-energy} (d). While the former costs about $91D$ to be created the latter needs an energy
about $123 D$. This process can be repeated in order to separate type 4 vertices
and the energy of each configuration can be readily obtained, in such a way that we can get the
potential energy of the system as a function of the distance between the vertices. Therefore, we have
used the same methods of Refs.~\onlinecite{Mol09,Mol10} to determine the potential that better describes
the interaction between type 4 vertices. Our results are very similar to those obtained in
these references, such that the potential reads:
\begin{equation} \label{eq} V(R)=q_t/R+b_{t}X(R) + c_{t}, \end{equation}
where $R$ is the distance between the monopoles and $X(R)$ is the string length.

\begin{figure}
\includegraphics[width=7cm]{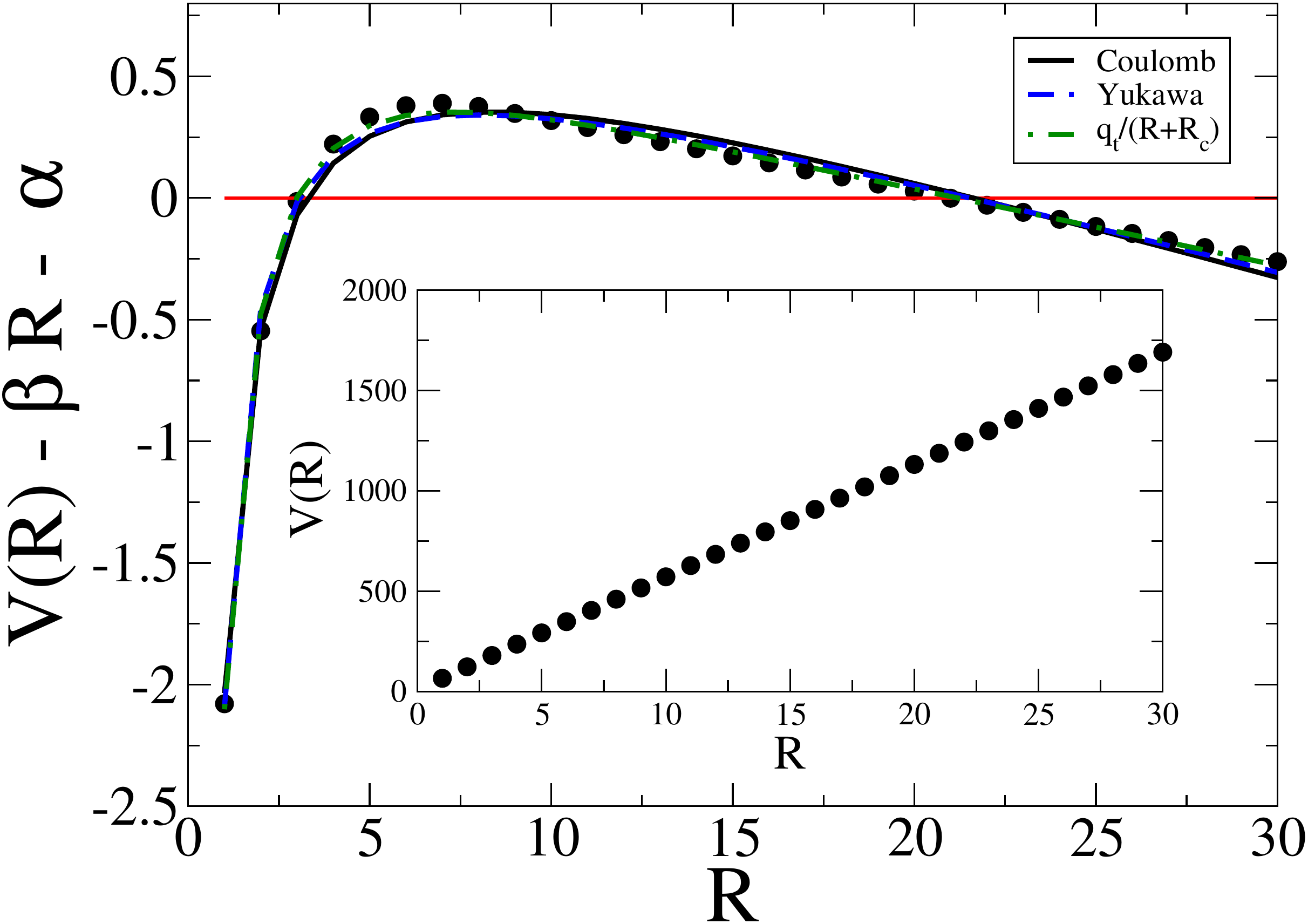}
\caption{ \label{pot} (Color online) Inset: Potential energy as a function of the distance ($R$)
between two single charged monopoles separated by a straight string. Outset: The black dots represent the potential
minus the linear contribution obtained by fitting $\beta R+\alpha$ to it. The lines are the best non-linear curve
fits of the potential according to the functions ($q_t/R+b_tR+c_t$) Coulomb, ($q_t\exp(-\kappa R)/R+b_tR+c_t$) Yukawa and
 $q_t/(R+R_c)+b_tR+c_t $ minus $\beta R+\alpha$.}
\end{figure}

Before showing numerical results some points deserve further explanation.
At a first glance, the potential function seems to be linear (see the inset of fig.~\ref{pot}).
However, after subtracting from the potential the linear contribution (see fig.~\ref{pot}),
one can see that a diverging contribution to the potential for small enough
distances must be added to the potential. Among the many possible functions that could be used we found that
the addition of a Coulombian term ($q_t/R$) to the potential is the best way to describe the ASSI and ATSI data
using a function with only three parameters. For the sake of comparison the $\chi^2/dof$ is about 0.3 for the pure linear fit
and about $6\times 10^{-4}$ for the potential of  Eq.~\ref{eq}. Besides that, as can be seen in fig.~\ref{pot}
the potential of Eq.~\ref{eq} better describes the data.
As is well known, one should be very careful when doing a non-linear curve fitting, specially when the fitted function
has many parameters, such that the use of functions with more than 3 parameters was avoided.
This is the reason why we have opted to describe the excitations by means of Eq.~\ref{eq}.
We have also tested other functional forms instead of the Coulomb one,
specially the Yukawa potential ($q_t\exp(- \kappa R)/ R $) and the function $q_t/(R+R_c)$, both with four parameters.
Using the Yukawa potential we found approximately the same values for the constants ($q_t$, $b_t$
and $c_t$) and the $\chi^2/dof$ diminished by less than one order of magnitude. The use of the
$q_t/(R+R_c)$  diminished the $\chi^2/dof$ by about one order of magnitude. Nevertheless the fit
convergence is highly dependent on the initial values, such that for all purposes we are considering
Coulombian interactions between the monopoles. It is worthy noticing that the values shown
here are approximate results, that depends on the number of points used in the non-linear fit. Differences up to
$20\%$ in $q_t$ may be expected.

When single-charged monopoles (type 4 vertices) are separated in a straight line by a straight string,
composed by type 5 vertices only (see fig.~\ref{low-energy} (e)), the constants in the potential are $q_t^l=-3.5Da$,
$b_t^l=56D/a$ and $c_t^l=13D$.
The monopoles can also be separated in a straight line by creating
only type 2 vertices (see fig.~\ref{low-energy} (f)), such that the string
has a sawtooth shape. In this case, the constants are $q_t^s=-3.9Da$,
$b_t^s=25D/a$ and $c_t^s=46D$, where the string length $X$ is related to the
charges distance, $R$, by $X=2R$. The huge difference in the string tension in these
situations can be easily understood by recalling that type 2 vertices has less energy than
type 5 vertices. It is important to stress that the string in this system is energetic, observable, and unique, in the sense that it is composed by type 2 or type 5 vertices on a background
of type 1 vertices. Besides, the value of $q_t$ is almost the same of that found in the square
spin ice ($-4 Da \leq q_s \leq -3.4 Da$)~\cite{Mol10}, while the string tension is
about six times higher when the string is composed by type 5 vertices only and
about two times higher when the string is composed by type 2 vertices ($b_s \sim 10 D/a$).
The difference in the string tension for the shapes considered indicates that there is
a huge anisotropy in the system.

\section{Ground-state considerations}

From an experimental point of view, the achievement of the system's ground-state is an
important step toward the study of monopoles physics. Considering the ATSI we believe that
two different approaches can drive the system to its ground-state, or at least to an state
close enough to it. The first one is the same approach
used by Morgan {\it et al}~\cite{Morgan10} to study the ASSI. In their work they studied the as growth
system and found the appearance of ground-state domains with sparse excitations that follows Boltzmann
statistics. Considering, for instance, the same island sizes and lattice spacings used by Morgan {\it et al}~\cite{Morgan10}
we may expect the achievement of larger domains in the ATSI as compared to the ASSI since in the ATSI,
the internal magnetic fields should be stronger. This is by virtue of the 
presence of more islands in the same area of the material. However, due to experimental difficulties, 
namely the resolution of the lithography process, it may not be possible to use the same sizes and lattice spacings used by
Morgan {\it et al}~\cite{Morgan10}. However, a simpler approach can also be very efficient to drive
the system to its ground-state as will be discussed in what follows.

Apply a strong magnetic field in the $-x$ direction of the sample (see Fig.~\ref{array}),
so that the system is driven to the magnetized state shown on the right side of Fig.~\ref{array},
similarly to what happens to ASSI. However, the energetics of these magnetized configurations, and those deviating 
from them, are not equivalent in the ASSI and in the ATSI. While the magnetized state of the ASSI is a local minimum,
since it is composed by vertices that satisfies the ice rule and any spin flip would increase the system's
energy, in the ATSI the magnetized state is a kind of saddle point. The ATSI's magnetized state is composed
by type 5 vertices only. If a spin in the $x$ direction is flipped, two type 4 vertices are
created and since type 4 vertices has less energy than type 5 vertices the system's energy diminishes.
The first spin flipped in the $x$ direction would release an energy of about 45 D. However, if a spin in any other direction is flipped
the energy of the system increases since two type 6 vertices are created (remember that type 6
vertices is more energetic than type 5 vertices). If the first spin flipped is not in the $x$ direction,
the system's energy increases by about 10 D. We may thus expect that, after applying a strong field in the $-x$ direction
another field applied
in the $+x$ direction would induce the formation of type 4 vertices. Once formed, these vertices can be
separated apart by creation of type 1 vertices reducing even more the system's energy.
In this process only spins in the $x$ direction (the field
direction) are expected to flip. This process would occur in such a way that monopole-like excitations are created
above a given state (which is not the ground-state, such that the monopole terminology does not seems to
be appropriate) and induced by the presence of the magnetic field to move along the field direction.
Note however that south poles (with four-out/two-in) in this situation move in the opposite direction to the field
as can be expected by analogy with basic electrostatics. If we restrict our analysis to vertices  energy only, we may see that spins that
are not in the $x$ direction are not expected to flip, at all, since the vertices that
would be created in such a process (type 6)  are much more energetic than the other vertices
present in the system (types 1, 4 and 5). Even in the presence of a relatively strong disorder
the energy needed to flip a spin that is not in the $x$ direction seems to be high enough
to prevent its occurrence. However, a detailed analysis of this hypothesis is required due to the long-range
character of dipolar interactions.
A complete treatment of this possibility is currently in progress and will be published elsewhere.

\section{\label{termo} Thermodynamics}

Knowing the ground state and the elementary excitations, the thermodynamics of the proposed system
should deserve attention. Despite the fact that in the artificial spin ices the moment configuration
is athermal, some experimental works have presented alternative methods to overcome this
difficulty \cite{Morgan10,Kapaklis11}. Materials with an ordering temperature
near room temperature \cite{Kapaklis11} are the most recent option; reduction
in island's volume and moment through state-of-the-art nanofabrication is another,
but it was not accomplished so far. Besides that, an effective
thermodynamic behavior can be obtained by using a rotating magnetic field\cite{Ke08,Nisoli07,Nisoli10}.
Independently of the experimental difficulties we have theoretically studied
some thermodynamic properties of the proposed array. Indeed, in an earlier paper considering 
the ASSI we have argued that the string should loose its tension due to entropic
effects \cite{Mol09} (the string configurational entropy), rendering
free monopoles at a temperature above a critical one, $T_{c}$. The same
arguments should be valid here, such that the verification
of this hypothesis is of great interest.

To study the thermodynamic properties of this system we have used conventional Monte Carlo techniques, which are
briefly presented. We have simulated lattices with size $L\times L_y$,
where $L_y$ is the largest integer resulting from $L/\sqrt{3}$, so that we have
$2\times L \times Ly$ vertices and $N_s=6\times L \times Ly$ spins for a given $L$;
we have considered systems with $L=16,24,32,48,64$ and $80$.
Although it is known~\cite{Mol11} that the introduction of a cut-off radius
in the evaluation of dipolar interactions may lead to a critical behavior that
does not agree with that found when full long-range interactions are considered,
we have opted in this work to use a cut-off radius at eight lattice spacings.
This choice is justified by our interest in the basic thermodynamic behavior
of the system and not on the details of the possible phase transitions. In our Monte
Carlo scheme a combination of single spin flips and loop moves were used. A loop move~\cite{Barkema98}
is a random closed path of aligned spins that are flipped or not according to
the Metropolis prescription ($p=\exp{(-\Delta E/k_BT)}$). One Monte Carlo step (MCS)
is the combination of $N_s$ single spin flips and $3$ loop moves. In most simulations
we have used $2\times 10^4$ MCS to equilibrate the system and $2\times 10^5$ MCS to
take averages.

\begin{figure}
\includegraphics[angle=0.0,width=7cm]{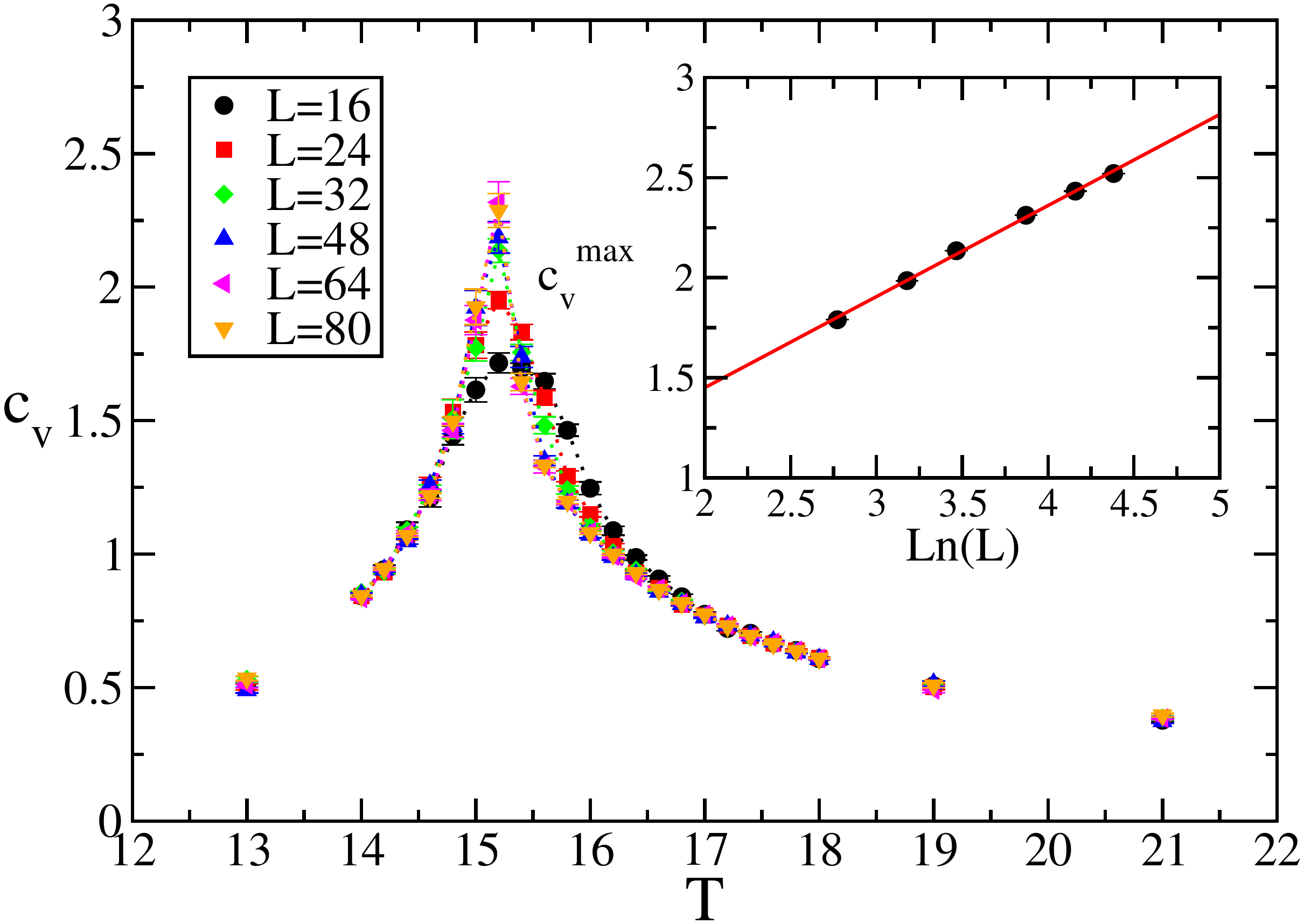}
\caption{ \label{SH} (Color online) Specific heat as a function of temperature for different lattice sizes. The inset shows the specific
heat maxima as a function of $\ln(L)$. The line is a guide to the eyes.}
\end{figure}

\begin{figure}
\includegraphics[angle=0.0,width=7cm]{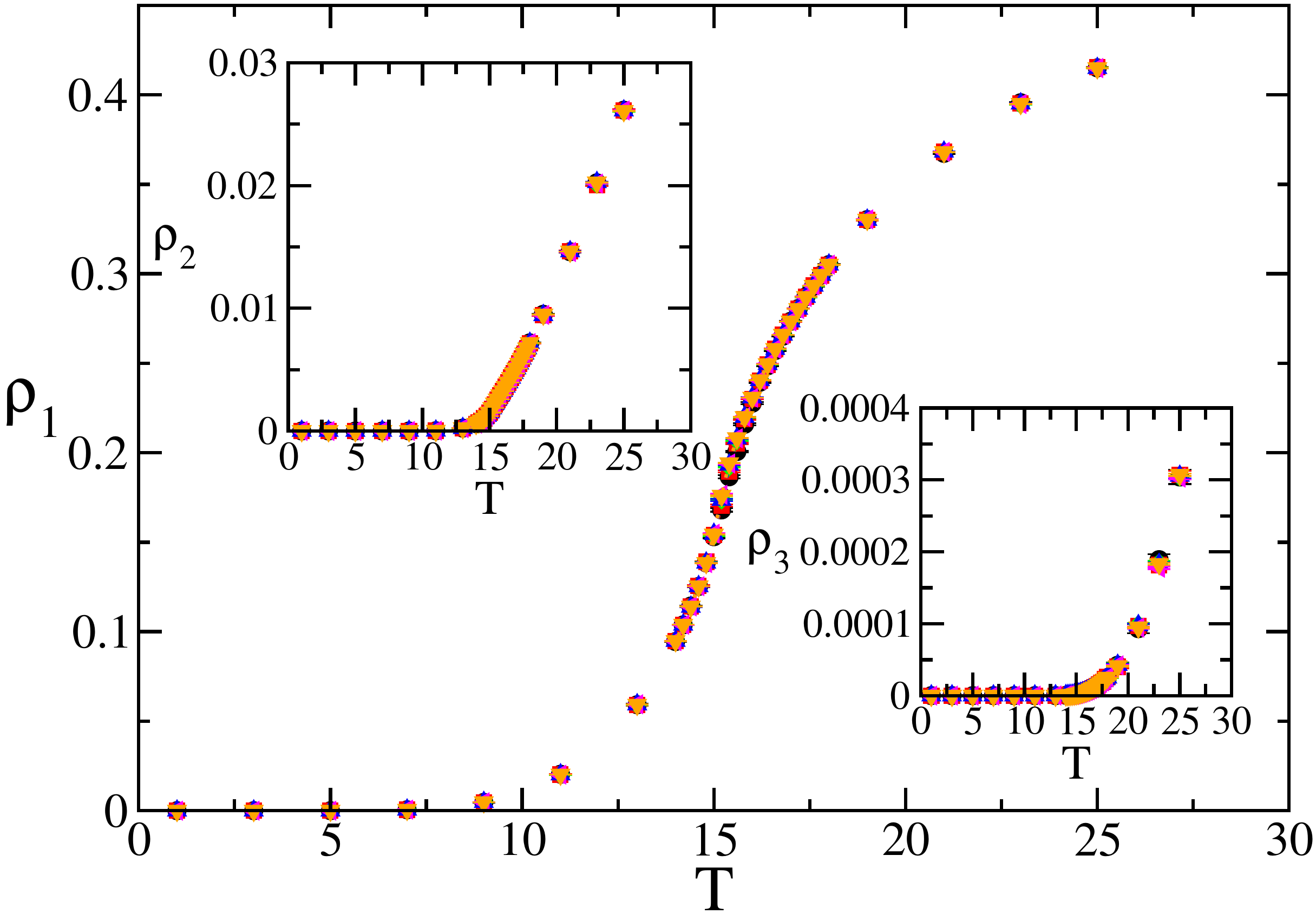}
\caption{ \label{Density} (Color online) Monopoles density as a function of temperature for different lattice sizes.
The outer plot shows the density of single charges ($\rho_1$) while in the insets the density of double ($\rho_2$)
and triple ($\rho_3$) charges are shown. The symbols are the same used in Fig.~\ref{SH}.}
\end{figure}

We start by presenting the results for the specific heat (see Fig.\ref{SH}). We notice that the specific
heat exhibits a peak at a temperature $T_{c}$ approximately equal to $15 D/k_B$.
The amplitude of this peak increases as the system size $L$ increases, as can be
seen in the inset of the same figure, where the specific heat maximum is plotted as
a function of $\ln(L)$. Therefore, it may indicate a
phase transition in the Ising universality class ~\cite{Privman}. We have also analyzed
the charges density and the average separation between monopoles and antimonopoles
as a function of $T$. In Fig.\ref{Density} we show the density of single charges (vertices types 3, 4 and 6),
$\rho_1$, double charges (type 7 vertices), $\rho_2$, and
triple charges (type 8 vertices), $\rho_3$, as a function of temperature. It is remarkable
that the density of double and triple charges are much smaller than the
density of single charges.

\begin{figure}
\includegraphics[angle=0.0,width=7cm]{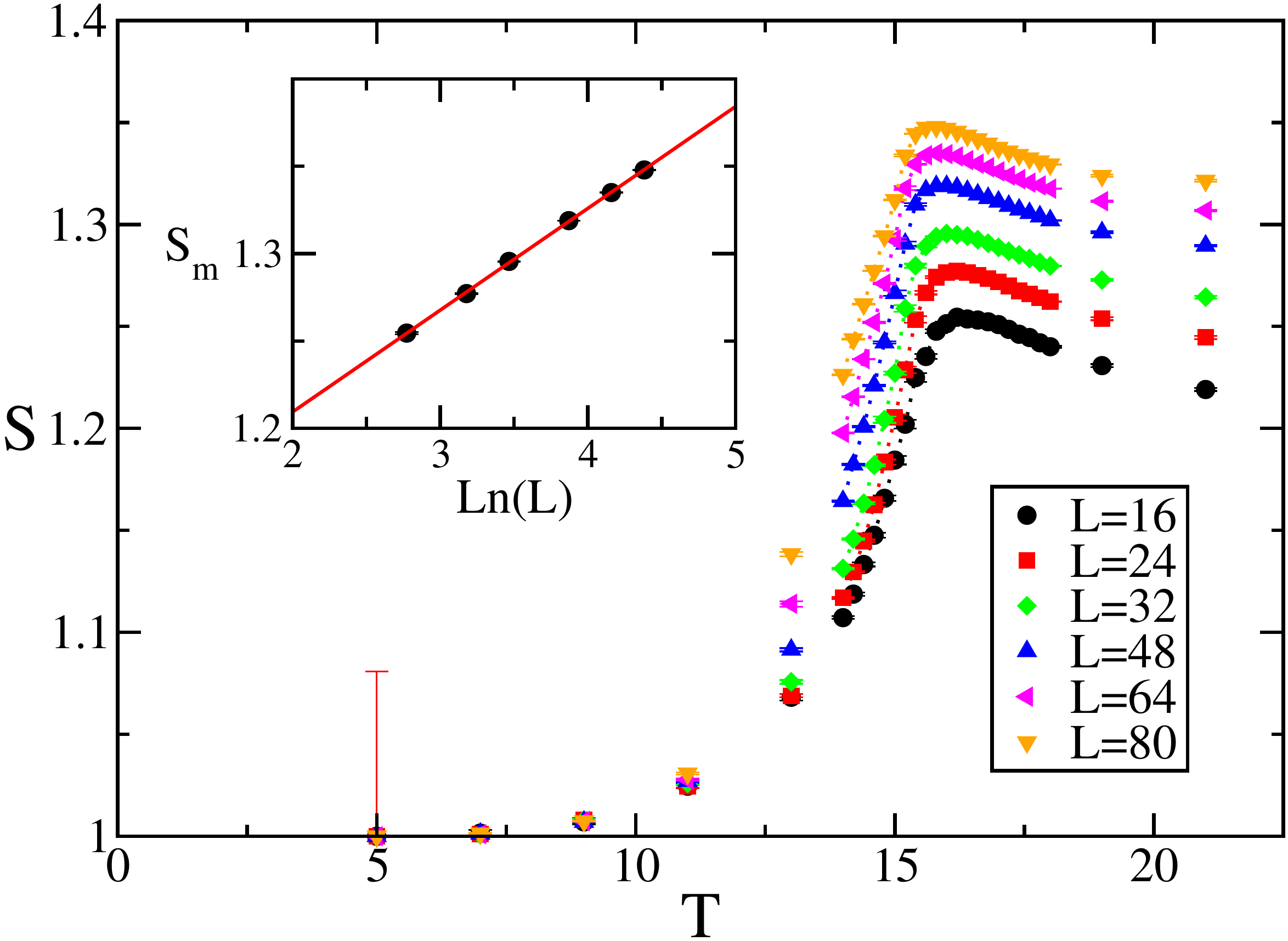}
\caption{ \label{Separation} (Color online) Mean distance between single charges as a function of temperature.
The inset shows the maximum mean distance as a function of $\ln(L)$. The line is a guide to the eyes.}
\end{figure}

The mean distance between monopoles and antimonopoles, $S$,
as a function of temperature was obtained by considering only defects
with unit charge. This quantity is calculated by using the method of assignment problems ~\cite{Bukard}.
In our case, we would like to assign $n$ positive charges to $n$ negative charges for a given configuration in
such a way that the sum of distances of all possible pairing be a minimum.
Note that studying the system's thermodynamics we are not able to determine the exact
pairs energy neither which charge forms a pair with which.
The results are shown in Fig.\ref{Separation}. The average separation has
a local maximum at the same temperature $T_{c}$ in which the specific heat
exhibits a peak ($ \sim 15D$). We notice that the amplitude of this
maximum also increases as the system size increases. Indeed, the maximum
average separation $S_{m}$ increases  logarithmically with the system size
$L$ ($S_{m} \propto \ln L$, see the inset of Fig.\ref{Separation}) and hence, one could
expect that a certain quantity of monopoles may be almost isolated for very
large arrays. For low temperatures the average separation $S$
does not depends on the lattice size $L$, while for high temperatures, this quantity has a tiny dependence
on $L$. This picture is suggestive and indicate a different behavior of monopoles and
strings at low and high temperatures \cite{Mol09,Silva12}.

At this point we should stress that the similarities between these results and the results
for the ASSI of Ref.~\cite{Silva12} are remarkable. As discussed before one would expect
many differences between those systems since some vertices that does not satisfy the ice
rule have less energy than some vertices that satisfy it. Indeed, the fact that the
some of the vertices of a string (type 5) have more energy than two kinds of charges (vertices types 3 and 4)
would be expected to considerably change the system's properties, specially those concerning
monopoles behavior. However, the only difference that can be
noticed is in the critical temperature, which is about $15D/k_B$ for the ATSI and $7D/k_B$
for the ASSI. The specific heat and magnetic charge behaviors are qualitatively the same, including
the mean distance between opposite magnetic charges and their finite size behavior. Of course, as the
string in the ATSI has stronger tension, the maximum value of the mean distance between the monopoles
is smaller than that of the ASSI for the same lattice size.

\section{Conclusions}

In summary, we have proposed and theoretically studied a realization of an artificial
spin ice in a triangular geometry. The usual experimental difficulty to achieve
the ground-state of the artificial square spin ice (ASSI) should not be so dramatic
for the artificial triangular spin ice (ATSI) array since a simple demagnetization
protocol (similar to a hysteresis loop) will probably drive the system to its ground-state
as discussed in section IV (see Fig.~\ref{array}).
This hypothesis is under current investigation.
Although there are fundamental differences between the ASSI and the ATSI, specially
concerning the vertices topologies, these systems show remarkable similarities. Indeed,
the lowest energy excitations of both can be described by magnetic monopoles excitations
interacting via a Coulombic term added by an energetic string potential; the thermodynamic
behaviors are qualitatively the same. However, contrary to the ASSI, in the ATSI there are 3 different
classes of magnetic monopoles, and much more work has to be done to completely understand if all kinds
of charged vertices behave in the same way. The same applies for the different classes
of three-in/three-out vertices.

Another issue that deserves further investigation is the similarity found in the monopoles' charge
value. While the string tension of the ATSI can be about six times larger than that of
the ASSI, the monopoles' charge have roughly the same value. Indeed, in Ref.~\onlinecite{Mol10}
we have shown that in a modified ASSI, where a height off-set is introduced between
the islands in the two different lattice directions, the string tension can be reduced
by a factor of about 20 and the monopoles' charge does not change appreciably.
For example, the smallest value of the string tension found for the modified ASSI
is about $0.59D/a$ and the highest value we found in this study is $56D/a$
which gives a difference of about 2 orders of magnitude. In general, 
we expect that the value of the string tension $b$ will not alter the physical 
picture of artificial systems with different geometries; it must only change the 
thermodynamic quantities, and $T_{c}$ should increase as $b$ increases (for instance, 
square and triangular lattices present similar thermodynamic properties but $T_{c}$ 
is lower in the square lattice, which has smaller $b$).
Remember that following our arguments on energy-entropy balance in 
ASSI~\cite{Mol09} we found that the effective string tension should be
given by $b-\epsilon k_BT$, with $\epsilon=\mathcal{O}(a^{-1})$. In this
way the critical temperature, the temperature at which the string looses its
effective tension, is proportional to the string tension $b$. Of course, $T_{c}\rightarrow 0$ 
if $b\rightarrow 0$ and hence, magnetic monopoles would be found free for any temperature. 
Besides, it is expected that smaller string tension demands a weaker field to
move the monopoles. On the other hand, all monopoles' charge values are in the range $-4 Da \leq q 
\leq -3.4 Da$~\cite{Mol10}. This observation may indicate that these charges have
a kind of universal behavior while the
string tension can be tuned by modifying the system's geometry. 
In addition, it is expected that whenever controlling the monopoles
motion, spin ice systems could be used for several practical applications. Indeed, the 
ability of designing systems with desired string tension, concerned
to the field necessary to move a charge, would benefit the development of new devices.
The main challenge should be then to find geometries in two-dimensions (or even 
in three dimensions ~\cite{Moller09,Mol10}) in which $b \sim 0$ making the
control of monopole excitations experimentally feasible by means of externally applied magnetic
fields.
Of course, a great deal of efforts must be done in order to verify the validity and applicability of these assumptions.

Another prospect for future investigation is the searching for a three-dimensional natural
version of this system, a hypothetical
``bulk spin ice" with three-in/three-out ice rule. In this case,
the expected crystalline structure would be
composed by Ising-like spins (as in Dy$_2$Ti$_2$O$_7$ and Ho$_2$Ti$_2$O$_7$) disposed in a network
of linked octahedra (which represents the central intersection of two tetrahedra) in such a way
that they are constrained to point along the line joining the center of adjacent octahedra.

The authors thank CNPq, FAPEMIG, FUNARBE and CAPES (Brazilian agencies) for
financial support.


\end{document}